\documentclass[twocolumn]{aastex631}
\pdfoutput=1
\usepackage[utf8]{inputenc}
\usepackage[T1]{fontenc}
\usepackage{nomencl}
\makenomenclature
\usepackage{amsmath}
\usepackage{natbib}
\usepackage{lmodern}
\usepackage{graphicx}
\usepackage{floatrow}%
\usepackage{lmodern}
\usepackage{booktabs}
\usepackage{float}
\floatstyle{plaintop}
\restylefloat{table}
\usepackage{etoolbox}
\robustify\itshape
\robustify\bfseries
\usepackage[caption=false]{subfig}
\DeclareCaptionLabelSeparator{emdash}{\textemdash}

\usepackage{savesym}
\savesymbol{tablenum}
\usepackage{siunitx}
\restoresymbol{SIX}{tablenum}

\accepted{Febraury 18, 2023}
\submitjournal{ApJ}

\shorttitle{X-ray spectroscopy of the PWN in HESS J1640-465 }
\shortauthors{Abdelmaguid et al.}

\begin{document}

\title{Broadband X-ray Spectroscopy of the Pulsar Wind Nebula in HESS J1640-465}

\author{Moaz Abdelmaguid}
\affiliation{New York University Abu Dhabi, PO Box 129188, Abu Dhabi, United Arab Emirates}
\affiliation{Center for Astro, Particle and Planetary Physics (CAP$^3$), New York University Abu Dhabi}
\author{Joseph D. Gelfand}
\affiliation{New York University Abu Dhabi, PO Box 129188, Abu Dhabi, United Arab Emirates}
\affiliation{Center for Astro, Particle and Planetary Physics (CAP$^3$), New York University Abu Dhabi}
\author{Eric Gotthelf}
\affiliation{Columbia Astrophysics Laboratory, Columbia University, 550 West 120th Street, New York, NY 10027-6601, USA}
\author{Samayra Straal}
\affiliation{New York University Abu Dhabi, PO Box 129188, Abu Dhabi, United Arab Emirates}
\affiliation{Center for Astro, Particle and Planetary Physics (CAP$^3$), New York University Abu Dhabi}

\begin{abstract}

We present updated measurements of the X-ray properties of the pulsar wind nebula associated with the TeV $\gamma$-ray source HESS J1640-465 derived from \emph{Chandra} \& \emph{NuSTAR} data. We report a high $N_{H}$ value along line of sight, consistent with previous work, which led us to incorporate effects of dust scattering in our spectral analysis. Due to uncertainties in the dust scattering, we report a range of values for the PWN properties (photon index and un-absorbed flux). In addition, we fit the broadband spectrum of this source and found evidence for spectral softening and decreasing unabsorbed flux as we go to higher photon energies. We then used a one zone time dependant evolutionary model to reproduce the dynamical and multi-wavelength spectral properties of our source. Our model suggests a short spin-down time scale, a relatively higher than average magnetized pulsar wind, a strong PWN magnetic field and maximum electron energy up to PeV, suggesting HESS J1640-465 could be a PeVatron candidate.

\end{abstract}

\keywords{supernova remnants,  neutron stars,  pulsars, pulsar wind nebula}

\section{Introduction} \label{sec:intro}

Since neutron stars are rapidly rotating and highly energetic, they power a highly relativistic electron positron wind which in turn interacts with the surrounding environment as it expands outwards creating a pulsar wind nebula (PWN). PWNe emit electromagnetic radiations through synchrotron emission in a wide range of energy going from radio up to X-rays due to the presence of highly relativistic electrons. In addition, the relativistic electrons can cause highly energetic emissions in the $\gamma$-ray part of the spectrum through inverse Compton scattering of ambient photons. 

HESS J1640-465 was first discovered by the HESS collaboration in 2006 \citep{2006ApJ...636..777A}. It was found to be spatially coincident with G338.3-0.0, a shell like supernova remnant (SNR)  previously identified through radio observations to have an 8$'$ diameter \citep{1996A&AS..118..329W}. The source was analyzed by \citet{2006ApJ...651..190L} using data obtained by Swift X-ray telescope and a possible connection with pulsar wind nebula was suggested. Later, \citet{2007ApJ...662..517F} reported the detection of a highly absorbed X-ray point source with extended emission in the field of view of HESS J1640-465 using XMM-Newton observations. However, due to low angular resolution, neither a pulsar nor the associated PWN were spatially resolved. \citet{2009ApJ...706.1269L} then managed to spatially resolve the pulsar and its associated PWN using observations from \emph{Chandra}. No X-ray emission has been detected from the SNR, likely due to its low temperature and the high value of interstellar absorption along the line of sight ($N_H \sim 1.4 \times 10^{23}{\rm cm}^{-2}$) \citep{2014ApJ...788..155G}. Using the 21 {\rm cm} HI absorption line, a distance of 8$-$13 kpc was derived for the source \citep{2009ApJ...706.1269L} which makes HESS J1640-465 the most luminious TeV source in our galaxy. \citet{2011A&A...536A..98C} derived upper limits on the radio flux with no detection of a radio counterpart and as of now, there is no $\gamma$-ray or radio pulsation from the source. Subsequently, \citet{2014ApJ...788..155G} reported a discovery of X-ray pulsations originating from a 206 ms pulsar (PSR J1640-4631) using data from NuSTAR as part of a survey covering the Norma Arm region of the galactic plane

\begin{table}[htbp]
\caption{Log of NuSTAR Observations of HESS J1640-465\label{tab:NuSTAR log}}
\makebox[\textwidth][c]{
\begin{tabular}{ccc}
\toprule
\toprule
Obs.ID  & Obs.Date  & Exposure (s) \\
\bottomrule
30002021002 & 2013 Sep 29  & 63920  \\
30002021003 & 2013 Sep 30  & 21356  \\
30002021005 & 2014 Feb 02  & 101758 \\
30002021007 & 2014 March 06  & 36986 \\
30002021009 & 2014 March 14  & 33692 \\
30002021011 & 2014 April 11  & 22540 \\
30002021013 & 2014 May 25  & 22197 \\
30002021015 & 2014 June 23  & 19679 \\
30002021017 & 2014 June 25  & 22463 \\
30002021019 & 2014 June 27  & 19468 \\
30002021021 & 2014 June 30  & 19799 \\
30002021023 & 2014 July 11  & 22549  \\
30002021025 & 2014 Aug 10  & 21941  \\
30002021027 & 2014 Sep 11  & 20351  \\
30002021029 & 2014 Oct 11  & 22218  \\
30002021031 & 2014 Nov 05  & 4570  \\
30002021033 & 2015 Jan 08  & 4419  \\
30002021034 & 2015 Jan 12  & 16774  \\
30002021036 & 2015 Feb 14  & 32513  \\
30002021038 & 2015 March 12 & 32089  \\
30002021040 & 2015 May 14 & 52328  \\
30002021042 & 2015 June 29 & 49137  \\
30002021044 & 2015 Aug 20 & 45892  \\
30002021046 & 2015 Oct 3 & 50346  \\
\bottomrule
\end{tabular}%
}
\label{tab:NuSTAR log}%
\end{table}%

\citep{2013ApJ...770..103H}. The pulsar was found to be spatially coincident with HESS J1640-465. It has a rapid spin-down rate of  $\dot{P} = 9.8 \times 10^{-13}$, a spin-down luminosity of $\dot{E} = 4.4 \times 10^{36}$ erg $s^{-1}$ and a characteristic age of $\tau_{c}$ = 3350 yr. \citet{2016ApJ...819L..16A} then performed a 2.3 year phase-connected timing analysis on this source and calculated its braking index to be p = $3.15 \pm{0.03}$ making it to be the first pulsar with a braking index $>$ 3.

\citet{2010ApJ...720..266S} reported the discovery of the Fermi source 1FGL J1640.8-4631 that is spatially coincident with HESS J1640-465. They concluded that the observed TeV emission arises from the PWN through inverse Compton Scattering. On the contrary, \citet{2014PhRvD..90l2007A} proposed that the observed emission would arise due to proton-proton interactions in the northern part of the spatially coincident SNR G338.4+0.1. \citet{2021ApJ...912..158M} conducted morphological analysis of the source using 8 years of Fermi-LAT data. This showed that HESS J1640-465 is largely extended in the GeV energy band. In addition to that, they modelled the broadband emission coming out of the source assuming two different scenarios, one where the emission is originating from the PWN linked to the pulsar PSR J1640-4631 and the second where the accelerated protons or electrons inside the SNR shock are responsible from the observed emission. For the first scenario, they managed to set an upper limit on the PWN's magnetic field as well as put some constrains on the spin-down power of the pulsar. As for the second scenario, the ratio of electrons to protons in the SNR shock must be high. And though the broadband emission could be reproduced under this scenario, they make the remark that it is less likely because of the large ratio between the radio upper limit and the $\gamma$-ray emission.

In \S \ref{sec:observations}, we outline the \emph{Chandra} and \emph{NuSTAR} observations used in this paper and describe our analysis steps and methods. In \S \ref{sec:spectral}, we report our X-ray spectral analysis results. And then in \S \ref{sec:modelling}, we apply an evolutionary model to multi-wavelength observations of HESS J1640-465 and then discuss the implications of our spectral energy distribution modelling in \S \ref{sec:discussion}. Finally, we summarize our findings and conclusions in \S \ref{sec:conclusion}.

\section{Observations and Data Reduction} \label{sec:observations}

\subsection{NuSTAR Observations and Data Reduction}\label{sec:nustar obs}

 The Nuclear Spectroscopic Telescope Array (\emph{NuSTAR}) is an X-ray facility operating between 3$-$79 keV. It has two co-aligned telescopes with detectors on their focal plane modules named FPMA and FPMB. Each telescope provides a $12'\times12'$ field of view with a point spread function (PSF) that has half power diameter (HPD) of $58''$ and full width half maximum (FWHM) of $18''$. HESS J1640-465 was observed by \emph{NuSTAR} over the years 2013, 2014 and 2015 for the purpose of following the timing evolution of the pulsar. This amounted to a total exposure time of $\sim$ 759 ks accumulated in 24 separate pointings. A log of these observations is presented in Table \ref{tab:NuSTAR log}.
 
\begin{figure}[ht!]
\includegraphics[scale=0.9]{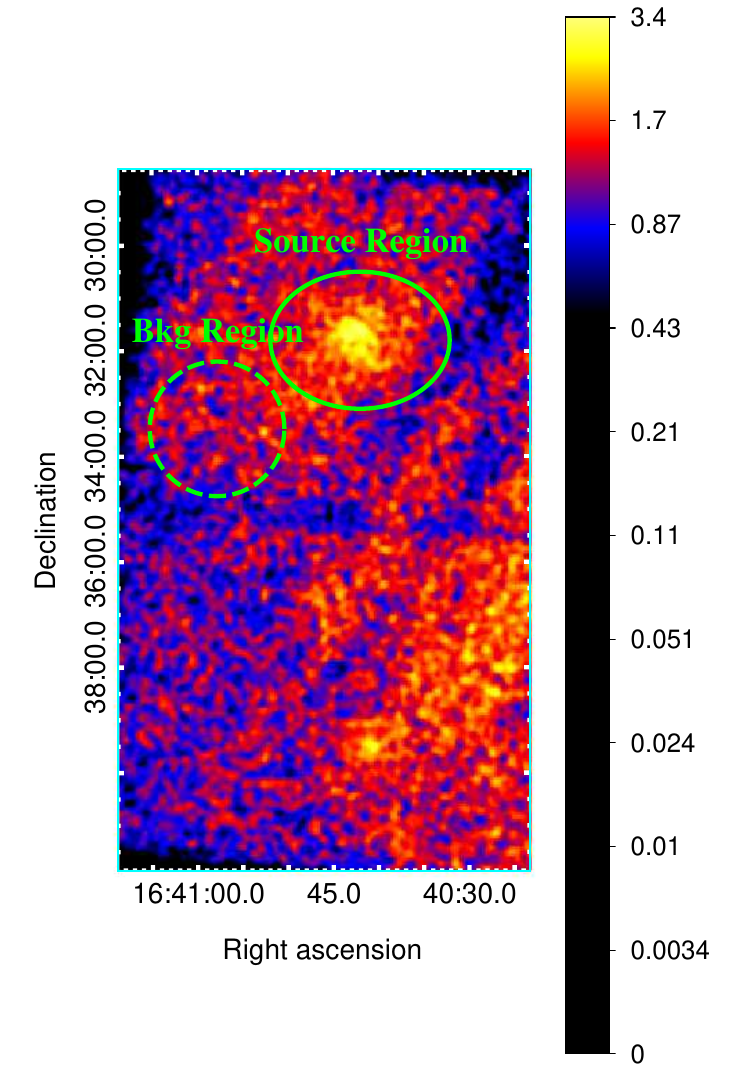}
\caption{NuSTAR image of HESS J1640-465 showing the regions used to extract the spectrum of the source and its associated background.}
\label{fig:nustar_selectiom}
\end{figure}

 Using  NuSTARDAS version 2.0.0 and HEASOFT version 6.28 available from the NASA High Energy Astrophysical Science Archive Research Center (HEASARC), we followed standard pipeline processing to generate cleaned and calibrated “Level 2 Data products” using the \textit{“nupipeline”} script. Once this stage was completed, we used the \textit{“nuproducts}” script with the option extended=yes to generate the associated response matrix (RMF) and ancillary response files (ARF) for an extended source.  
 
 We extracted individual spectra from the PWN using an elliptical region of diameter $3.4' \times 2.6'$ to ensure consistency with \citet{2014ApJ...788..155G}. We also used an offset circular region of radius $76.6''$ to account for background emissions. Figure \ref{fig:nustar_selectiom} shows the source and background regions used for extraction. Spectra from each focal module were combined as well as their respective response files/matrices using the \emph{addascaspec} routine in XSPEC version 12.11.1 \citep{1996ASPC..101...17A} and subsequent fitting was also performed in XSPEC.
 
\subsection{Chandra Observations and Data Reduction}\label{sec:chandra obs}

HESS J1640-465 has been observed using \emph{Chandra's} Advanced CCD Imaging Spectrometer (ACIS)\citep{2003SPIE.4851...28G}. Each ACIS chip detects photons in the energy range of 0.2--10 keV with a field of view 8$'$.3 $ \times$ 8$'$.3 and imaging resolution of $\sim$ 1$''$. For our analysis, we analyze ACIS-I observations taken in 2015 with a total observation time of $\sim$ 150 ks, as detailed in Table \ref{tab:Chandra log}. All Observations were in timed exposure and very faint mode. Obs.Id 17582 was removed from our analysis due to its short exposure time $(\sim$9 ks). This data-set comprises $\sim$ 3x more exposure time than that analyzed in previous work by  \citet{2014ApJ...788..155G}.

\begin{table}[htbp]
\caption{Log of Chandra Observations of HESS J1640-465 analyzed in this paper}
\makebox[\textwidth][c]{
\begin{tabular}{ccc}
\toprule
\toprule
Obs.ID  & Obs.Date  & Effective Exposure \\
&YYYY MM DD&(Ks) \\
\bottomrule
16771 & 2015 Jan 22 &   39.54 \\
17579 & 2015 Jan 23 &   34.60  \\
17580 & 2015 Jan 24 &   30.33 \\
17581 & 2015 Jan 25 &   43.82\\
\bottomrule
\end{tabular}%
}
\label{tab:chandra_log}%
\end{table}%

\begin{figure}[htbp!]
\includegraphics[scale=0.9]{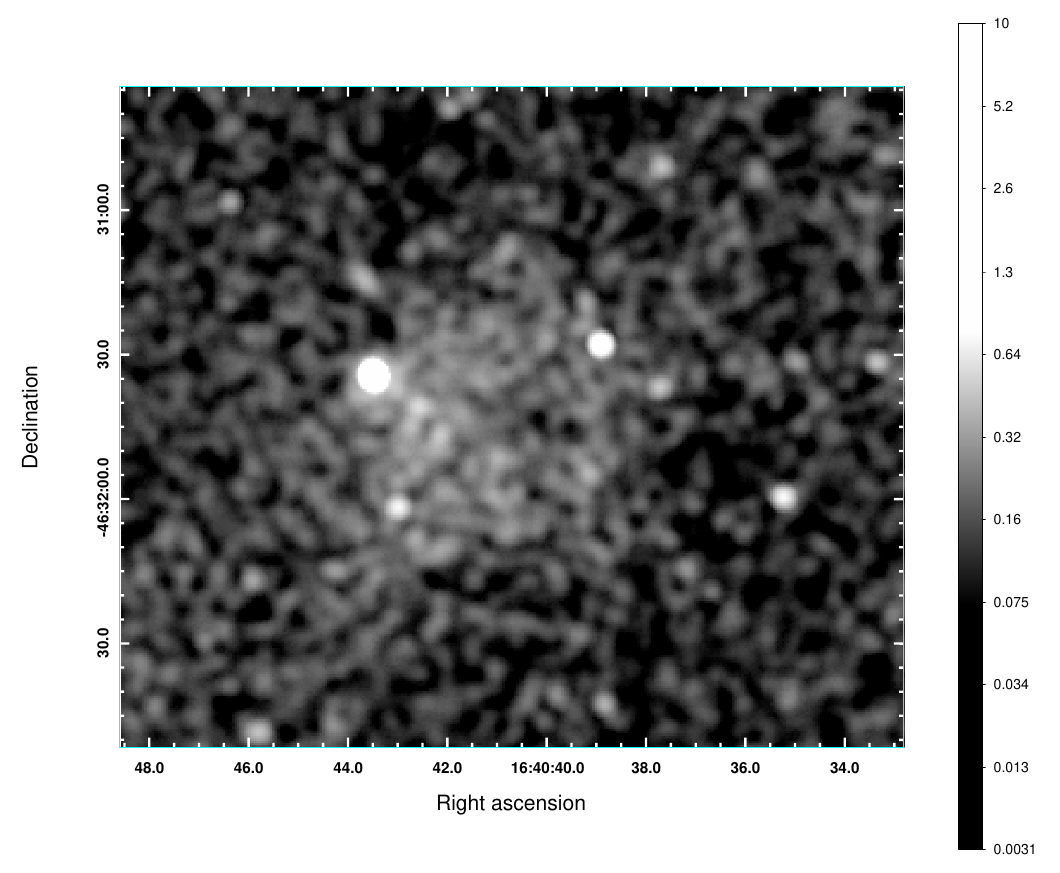}
\caption{Combined Chandra image showing the pulsar and its extended diffuse emission from the PWN. The image is smoothed to highlight the emission from the PWN.} 
\label{fig:source}
\end{figure}

Each observation was reprocessed individually using the $chandra\_repro$ package of the \emph{Chandra} Interactive Analysis of Observations (CIAO) software suite version 4.13 \citep{2006SPIE.6270E..1VF}, which resulted in a reprocessed $event\_2$ file for each observation. The reprocessed event files were then merged using the $reproject\_obs$ routine to create a deeper combined image of the source. Figure \ref{fig:source} shows the resulting image of the source after combining the four data sets. This is a zoomed in, smoothed image of the source highlighting the region of the pulsar and its extended diffuse emission associated with the PWN. We extracted spectra from the individual observations logged in Table \ref{tab:chandra_log} along with their response matrices using CIAO's $specextract$ tool. The spectra were then co-added using the $combine\_spectra$ routine in Sherpa version 4.13.0 \citep{2001SPIE.4477...76F}.

\begin{figure}[htbp!]
\includegraphics[scale=0.82]{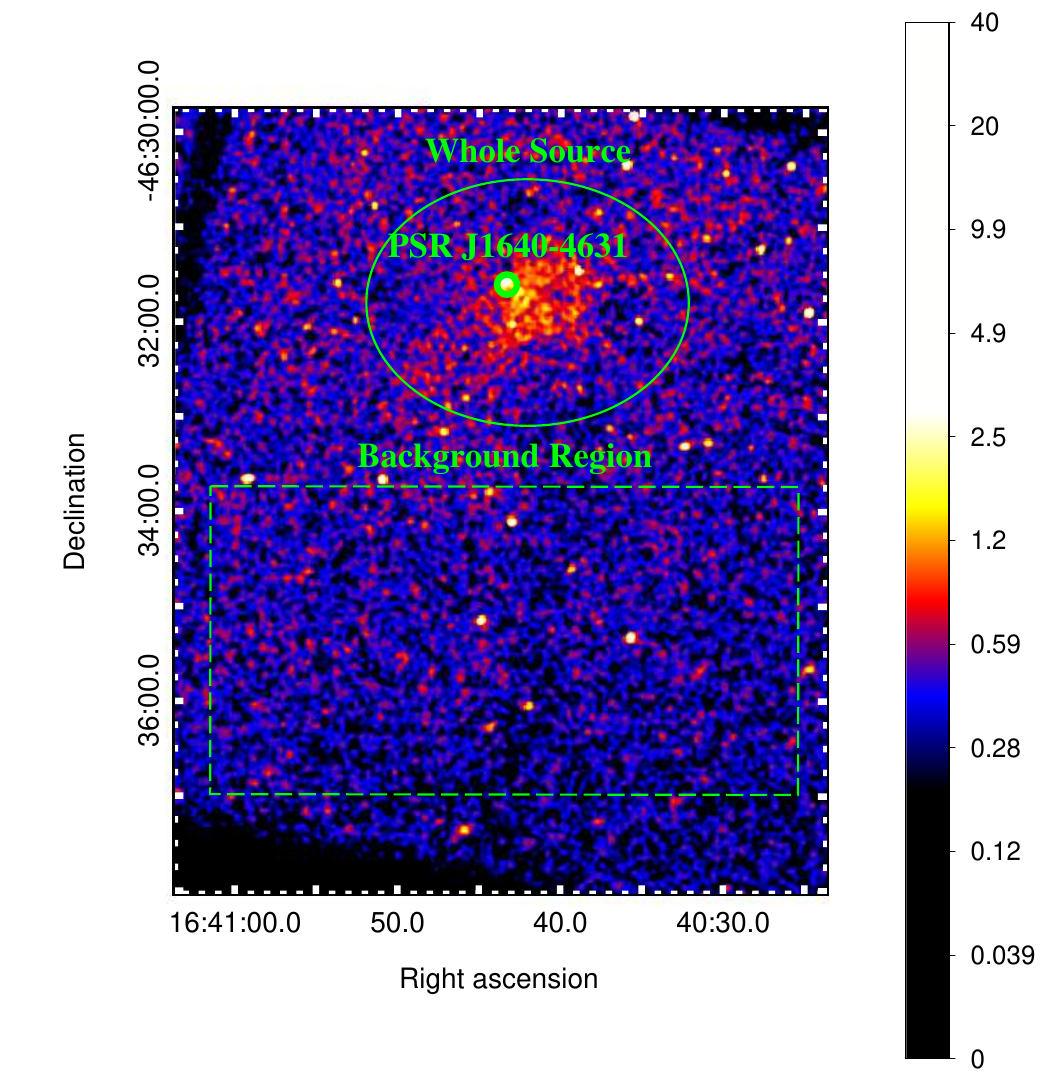}
\caption{Combined Chandra image showing the extraction region used for the whole source (Elliptical), PSR J1640-4631 only (Circular) and background region (Rectangular).} 
\label{fig:extraction}
\end{figure}

For the pulsar, we used a circular region of radius 2.8$''$ to extract the spectrum and its associated response files for a Point-like source using CIAO version 4.13 \citep{2006SPIE.6270E..1VF} with CALDB version 4.9.4. For the whole source, we used the same elliptical region we used earlier for \emph{NuSTAR} to account for scattering along line of sight as discussed in \S \ref{sec:scattering}. By using this large source region, we ensure all the scattered emission is taken into consideration. We extracted the spectrum and response files for an extended object also using CIAO version 4.13 \citep{2006SPIE.6270E..1VF}. We used a rectangular background region chosen to be far from the source region to minimize possible contamination due to extended emission coming from the PWN. Figure \ref{fig:extraction} shows the chosen extraction region for the whole source, pulsar only and their associated background region. The background subtracted total of 405 photons across all observations were detected from the pulsar. We investigated the possibility of pileup using the CIAO routine $pileup\_map$, but we found no evidence of pileup in the data. 

\section{Spectral Analysis} \label{sec:spectral}

In \emph{NuSTAR} observations, the pulsar and PWN are not spatially resolved. Only \emph{Chandra} has sufficient angular resolution to spatially resolve the emission coming out of the pulsar from that of the PWN. Therefore, to constrain the PWN spectrum in the data obtained by \emph{NuSTAR}, we will do the following: 
\begin{itemize}
    \item Quantify the flux systematic offsets (Cross calibration constant) between \emph{Chandra} and \emph{NuSTAR} as discussed in \S \ref{sec:cross_normalization}
    \item Analyze \emph{Chandra's} pulsar data alone to account for scattering along line of sight as discussed in \S \ref{sec:scattering} and \S \ref{sec:pulsar results}
    \item Use the un-scattered pulsar flux along with the cross normalization factor to jointly fit \emph{Chandra} and \emph{NuSTAR} spectra to obtain accurate measurement of the PWN spectrum as discussed in \S \ref{sec:nustar}
\end{itemize}

For spectral fitting, we used XSPEC version 12.11.1 \citep{1996ASPC..101...17A} and we accounted for the column density using the Tuebingen-Bolder ISM absorption model, implemented in Xspec as the \textbf{TBabs} with solar abundances set to wilms \citep{2000ApJ...542..914W}. The fit statistic was set to $\chi^{2}$ in all of our fits and XSPEC's Markov Chain Monte Carlo (MCMC) method was used to obtain the uncertainties in our model parameters and all the reported uncertainties in this paper are reported at the $90\%$ confidence level. 

\subsection{Cross Normalization/Calibration} \label{sec:cross_normalization}

Since we jointly fit data from two different observatories (\emph{Chandra} and \emph{NuSTAR}), cross calibration is important. To find the cross normalization between the two missions for our source, we extracted the \emph{Chandra} spectra for the whole source using the same elliptical region used to extract the \emph{NuSTAR} spectra. In doing that, emission from both the pulsar and PWN is included in our selected region. Figure \ref{fig:extraction} shows the extraction region used for the whole source used to find the cross normalization factor. 

For \emph{NuSTAR}, we initially fitted spectra from FPMA and FPMB independently and got consistent results. Therefore throughout this paper, we will be fitting them simultaneously. 

We first jointly fit the \emph{NuSTAR} and \emph{Chandra} spectra in their overlapping 3$-$10 keV energy range using an absorbed power-law model, $tbabs \times pegpwrlw \times constant$ (in XSPEC) with photon index $(\Gamma)$ and hydrogen column density ($N_H$) linked across the two data-sets. This constant was fixed to 1 for \emph{Chandra} and allowed to vary for \emph{NuSTAR} that resulted in a value of 0.75 for the \emph{NuSTAR} cross calibration constant, which is comparable to the value obtained by \citet{2017AJ....153....2M}. This value will be subsequently used for all our joint fits in this paper.

\subsection{Chandra Data} \label{sec:chandra_data}
For the pulsar, we used the circular extraction region shown in Figure \ref{fig:extraction}. We restricted our fit to be between 0.2$-$10 keV and grouped the pulsar spectrum to have 25 counts per bin and used an absorbed power law  ($tbabs \times pegpwrlw$) to fit the data \citep{2016ApJ...831..145Y}.

\subsubsection{Results} \label{sec:pulsar results}

\begin{table}[htbp]
\caption{Best fit spectral parameters for PSR J1640-4631 using data from Chandra}
\makebox[\textwidth][c]{
\begin{tabular}{ccc}
\toprule
\toprule
Parameter & Our Work  & \citep{2014ApJ...788..155G} \\
\bottomrule
$N_H({\rm cm}^{-2})$  &  ($1.4^{+0.5}_{-0.4}) \times 10^{23}$ & $(1.2\pm 0.6) \times 10^{23}$ \\ 
$\Gamma_{PSR}$ &  $0.6^{+0.6}_{-0.5}$ & $1.2^{+0.9}_{-0.8}$\\
$F_{PSR}$(2$-$10 keV) &  $ 2.5^{+0.4}_{-0.3} \times 10^{-13}$ &$1.9^{+0.2}_{-0.8} \times 10^{-13}$ \\
$\chi_{\nu}^{2}(DOF)$&  0.98 (18)& 1.0 (56)    \\ 
\bottomrule
\end{tabular}%
}
\tablecomments{ A comparison between best fit spectral parameters for the pulsar using an absorbed power law. This is using data from Chandra only. $F_{PSR}$ is the absorbed flux in the 2$-10$ keV. The reported fluxes are in terms of erg ${\rm cm}^{-2}s^{-1}$. Errors are at the 90 $\%$ confidence level.}
\label{tab: pulsar data alone}
\end{table}%

\begin{figure}[ht!]
\includegraphics[scale=0.5]{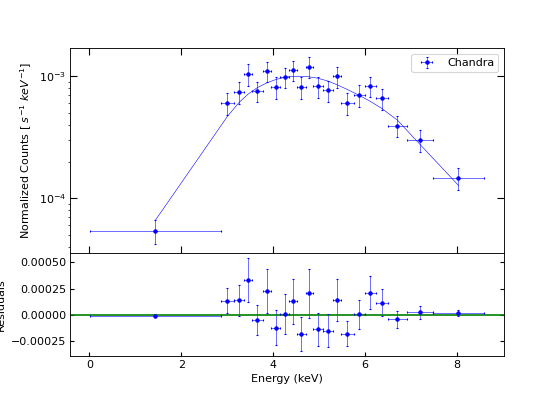}
\caption{Chandra's spectrum of PSR J1640-4631 fitted with an absorbed power-law reported in Table \ref{tab: pulsar data alone}. The lower panel shows residuals from the best fit.}
\label{fig:pulsar_fit}
\end{figure}

For the pulsar, we obtained a best fit $N_H = (1.4^{+0.5}_{-0.4}) \times 10^{23}$ $ {\rm cm}^{-2}$, $\Gamma = 0.6^{+0.6}_{-0.5}$ and an absorbed flux at 2$-$10 keV, F = $2.5^{+0.4}_{-0.3} \times 10^{-13}$ erg ${\rm cm}^{-2}$ $s^{-1}$. For this fit, the reduced $\chi^{2}_{\nu}$ = 0.98 for 18 degrees of freedom and the absorbed flux at 2$-$10 keV is reported for comparison with previous work done by \citet{2014ApJ...788..155G}. The resulting spectral fit is shown in Figure \ref{fig:pulsar_fit} and the best fit parameters are tabulated in Table \ref{tab: pulsar data alone} in which we compare our results with earlier work on this source done by \citet{2014ApJ...788..155G}.  While our results generally agree with \citet{2014ApJ...788..155G}, we set a tighter constraint on the pulsar properties due to the larger exposure time analyzed in this paper.

\begin{table*}[htbp]
\caption{Effects of scattering on the flux and photon indices values of \emph{Chandra's} pulsar}
\makebox[\textwidth][c]{
\begin{tabular}{cc|ccc|ccc}
\toprule
\multicolumn{2}{c}{Parameters}  & \multicolumn{3}{c}{Dust Model 1 (MRN)}  & \multicolumn{3}{c}{Dust Model 2 (ZDABAS) }\\
$N_H({\rm cm}^{-2})$& Distance & $\Gamma_{PSR}$ &   Flux &$\chi_{\nu}^{2}(DOF)$&$\Gamma_{PSR}$& Flux & $\chi_{\nu}^{2}(DOF)$ \\ \midrule
$1.0 \times 10^{23}$ & 0.1 &$1.4\pm{0.3}$ &$3.9\pm{0.4}$ &1.01 (19) &$1.2\pm{0.4}$ &$3.7\pm{0.4}$& 0.96 (19)\\ 
$1.0 \times 10^{23}$ & 0.3 & $1.3\pm{0.3}$& $3.8\pm{0.4}$& 0.98 (19)   &$1.1\pm{0.4}$&$3.6\pm{0.4}$&0.94 (19)\\ 
$1.0 \times 10^{23}$ & 0.5 & $1.1\pm{0.3}$&$3.6\pm{0.4}$  & 0.95 (19) &$0.9\pm{0.3}$ &$3.4\pm{0.4}$ & 0.94 (19) \\ 
$1.0 \times 10^{23}$ & 0.7 &$\mathbf{1.4\pm{0.3}}$&$\mathbf{3.9\pm{0.4}}$  &\bfseries{1.00 (19)} &$1.1\pm{0.3}$& $3.6\pm{0.4}$ & 0.94 (19)\\ 
$1.0 \times 10^{23}$ & 0.9&$0.9\pm{0.3}$&$3.5\pm{0.4}$ & 0.91 (19)  &$\mathbf{0.8\pm{0.3}}$&  $\mathbf{3.3\pm{0.4}}$& \bfseries {0.91 (19)} \\ 
\hline
\hline
$1.4 \times 10^{23}$ & 0.1 & $2.1\pm{0.4}$& $4.8\pm{0.4}$ &1.46 (19) &$1.8\pm{0.4}$&$4.3\pm{0.4}$ & 1.28 (19)\\
$1.4 \times 10^{23}$ & 0.3 & $1.9\pm{0.4}$ & $4.5\pm{0.4}$ &1.35 (19)&$1.7\pm{0.4}$&$4.1\pm{0.4}$ & 1.21 (19)\\
$1.4 \times 10^{23}$ & 0.5 & $1.6\pm{0.3}$ & $4.1\pm{0.4}$ & 1.20 (19) &$1.4\pm{0.4}$&$3.9\pm{0.4}$ & 1.10 (19) \\
$1.4 \times 10^{23}$ & 0.7 & $\mathbf{2.1\pm{0.4}}$&$\mathbf{4.9\pm{0.4}}$ & \bfseries{1.49 (19)} &$1.8\pm{0.4}$&$4.3\pm{0.4}$ & 1.29 (19)\\
$1.4 \times 10^{23}$ & 0.9 &$1.4\pm{0.4}$&$3.8\pm{0.4}$  & 1.10 (19) &$\mathbf{1.3\pm{0.4}}$ &$\mathbf{3.6\pm{0.4}}$ &\bfseries{ 1.03 (19)}  \\
\hline
\hline
$1.9 \times 10^{23}$ & 0.1 &{\itshape 2.9} & {\itshape 7.0}& $>$ {\itshape 2.0} &  $2.5\pm{0.4}$ &$5.6\pm{0.4}$ & 1.96 (19) \\
$1.9 \times 10^{23}$ & 0.3 & $\mathbf{2.7\pm{0.4}}$ &$\mathbf{6.0\pm{0.5}}$ &\bfseries{1.93 (19)}&   $2.4\pm{0.4}$&$5.2\pm{0.4}$ & 1.83 (19) \\
$1.9\times 10^{23}$ & 0.5 & $2.4\pm{0.4}$&$5.1\pm{0.4}$&1.81 (19) & $2.1\pm{0.4}$ &$4.6\pm{0.4}$& 1.62 (19)\\
$1.9 \times 10^{23}$ & 0.7 & \emph{2.9}& \emph{6.8} &  $>$  {\itshape 2.0} & $2.5\pm{0.4}$ &$5.5\pm{0.4}$ & 1.98 (19)\\
$1.9 \times 10^{23}$ & 0.9 & $2.1\pm{0.4}$ & $4.5\pm{0.4}$  & 1.64 (19) & $\mathbf{1.8\pm{0.4}}$& $\mathbf{4.1\pm{0.4}}$ & \bfseries{1.48 (19)} \\
\bottomrule
\end{tabular}%
}
\tablecomments{ The effect of scattering on the pulsar's flux and photon indices values. The $N_H$ values used in the scattering model are the 90 $\%$ limits we obtained earlier from our analysis of \emph{Chandra's} pulsar data. Distance refers to the dust position along the line of sight with 0 indicating dust grains located at the observer and 1 at the source.  The flux values are in terms of erg ${\rm cm}^{-2}s^{-1}$. Errors are at the 90 $\%$ confidence level.}
\label{tab:scatteringl}%
\end{table*}%

\begin{table*}[htbp]
\caption{Range in X-ray spectrum of pulsar derived from \emph{Chandra} data accounting for uncertainty in dust scattering along line of sight}
\makebox[\textwidth][c]{
\begin{tabular}{c|cc|cc|cc}
\toprule
\toprule
Parameter & \multicolumn{2}{c}{Low Boundary} &\multicolumn{2}{c}{Best Fit} &\multicolumn{2}{c}{High Boundary} \\
\bottomrule
 \multicolumn{7}{c}{Derived Pulsar Properties}\\
 & (1) & (2) & (3) & (4)& (5) & (6) \\
$N_H({\rm cm}^{-2})$ & $1.0 \times 10^{23}$  &  $1.0 \times 10^{23}$&  $1.4 \times 10^{23}$ &  $1.4 \times 10^{23}$&  $1.9 \times 10^{23}$& $1.9 \times 10^{23}$\\
$\Gamma_{PSR}$ &  0.8 & 1.4 & 1.3 &2.1 & 1.8&2.7  \\
$F_{PSR}$ (3$-$10 keV) &  $ 3.3 \times 10^{-13}$& $ 3.9 \times 10^{-13}$ &$ 3.6 \times 10^{-13}$ &$ 4.9 \times 10^{-13}$ &$ 4.1 \times 10^{-13}$& $6.0 \times 10^{-13}$ \\
$\chi_{\nu}^{2}(DOF)$ &0.91 (19) & 1.00 (19) & 1.03 (19) &1.49 (19) &1.48 (19) & 1.93 (19) \\
\bottomrule
\end{tabular}%
}
\tablecomments{The reported fluxes are un-absorbed and in the units of erg ${\rm cm}^{-2}$ $s^{-1}$.}
\label{tab:psr_properties}
\end{table*}%

\subsection{Scattering Effects} \label{sec:scattering}
As X-ray photons travel through the interstellar medium (ISM), they can be absorbed or scattered, which affects the measured spectrum within a particular source region around  astrophysical sources. The Scattering of X-ray photons by dust along the line of sight forms x-ray halo around these objects. As a result, the spectrum of photons detected in a small region around the object is both lower and harder than the intrinsic spectrum of the source \citep{2016ApJ...818..143S}. Since our source is highly absorbed ($N_H$ = $1.4 \times 10^{23}$ ${\rm cm} ^{-2}$), and consistent with earlier studies (e.g, \citealt{2010ApJ...720..266S} \&  \citealt{2014ApJ...788..155G} ), we need to take scattering into consideration in our spectral analysis.  We incorporate these effects  using the $xscat$ model in XSPEC \citep{2016ApJ...818..143S}. This model calculates the un-scattered intrinsic pulsar spectrum taking into account the fitted photon energy range, model for dust, extraction region and distance between the dust and the source. Since we have no prior information on either the type of dust or its location along the line of sight, we decided to explore the derived intrinsic values of pulsar photon indices and un-absorbed fluxes for different dust models over assumed range of $N_{H}$ values derived from our analysis of the pulsar data.  

For dust model, we used two of $xscat's$ models; first is the MRN model based on the work of \citet{1977ApJ...217..425M} and the second is the ZDABAS model based on the work of \citet{2004ApJS..152..211Z}, and we allow the relative location of dust grains along line of sight from the source to vary between 0 (the dust grains are at the source) and 1 (the dust grains are at the observer).

\subsubsection{Results of Scattering Effects} \label{sec:pulsar results}

 Table \ref{tab:scatteringl} shows our results when we include the $xscat$ model in our spectral fits. For the same value of $N_{H}$, the MRN and ZDABAS dust models resulted in significant effects on the pulsar spectrum depending on how far the dust is from our source. It is worth mentioning that the four italicized values without error bars in Table \ref{tab:scatteringl} in the MRN dust model resulted in a bad fit with $\chi_{\nu}^{2}$ $>$ 2, so we don't take them into consideration. In addition, at $N_{H} = 1.9 \times 10^{23} {\rm cm}$, our fits result in  higher $\chi_{\nu}^{2}$ values than at lower $N_{H}$ as detailed in Table \ref{tab:scatteringl}, which suggests that a lower column density value is favoured. However, we decided to include all results in our PWN analysis for completeness.

 For both dust models, we observe a similar trend, but systematic offsets in the two parameters we are observing. To be more specific, for the same value of $N_{H}$ at the same distance, the MRN model always results in higher flux value and softer (larger) photon index than the ZDABAS model. 

\begin{center}
\begin{figure*}
\subfloat[MRN Dust Model]{\label{fig:sub_1}
\includegraphics[width=0.48\linewidth]{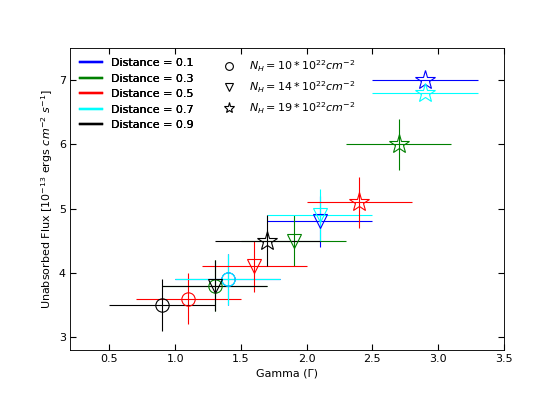}
}
\hfill
\subfloat[ZDABAS Dust Model]{\label{fig:cdiagram}
\includegraphics[width=0.48\linewidth]{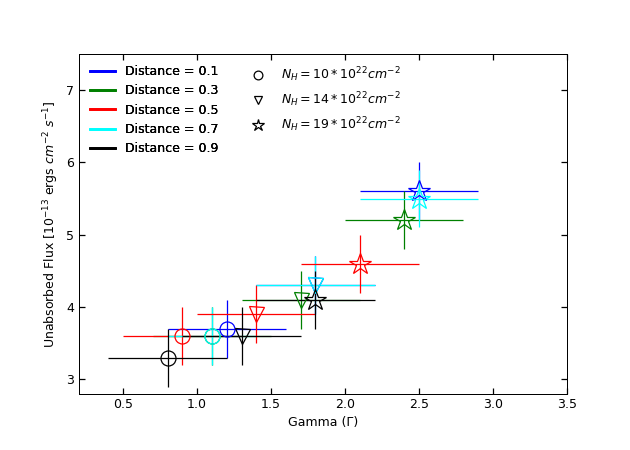}
}
\caption{Un-absorbed flux vs Gamma ($\Gamma$) of the pulsar at a range of locations of the dust grain along line of sight assuming our previously obtained $N_H$ values. Each colour represents a particular distance as indicated on the legend, and for the same distance (colour), each marker represents a different $N_H$ value. Same scale and axes limits were used for ease of comparison and visualisation. (a) MRN dust model. (b) ZDABAS dust model.}
\label{fig:scattering}
\end{figure*}
\end{center}

 For different values of $N_{H}$ at the same location of dust along line of sight, increase in $N_{H}$ results in an increase in flux and photon index. This trend is seen in both models as shown in Figure \ref{fig:scattering}. 

 To accommodate as large of an uncertainty range as we can and propagate that to the \emph{NuSTAR} data-set to obtain a more accurate PWN spectrum that will be used in the modelling later, we therefore selected two points for each $N_{H}$ value; one representing the lowest flux and hardest spectrum (lowest $\Gamma_{PSR}$) and the other representing the highest flux and softest spectrum (highest $\Gamma_{PSR}$). Table \ref{tab:psr_properties} shows our final derived values for the pulsar properties after taking dust scattering into consideration. In total, we have 6 different combinations of parameters that will be used in fitting the \emph{NuSTAR} data in the next section. 
  
\subsection{NuSTAR Data} \label{sec:nustar}
For our source, we used the extraction region shown in Figure \ref{fig:nustar_selectiom} and 
grouped the spectrum to have at least 5 $\sigma$ per spectral bin. Though \emph{NuSTAR} can detect photons with energies up to 79 keV, emission from our source is only brighter than the background below 30 keV, and therefore we only fit the measured spectrum below this energy as shown in the right panel of Figure \ref{fig:pwn_fit}.

As previously mentioned, \emph{NuSTAR} lacks sufficient angular resolution to spatially resolve the emission coming from the pulsar and the PWN. To overcome this, we will fit the composite spectrum with two absorbed power law components, one to account for emission from the pulsar and the other to account for emission from the PWN. The properties of the pulsar component will be fixed to the values in Table \ref{tab:psr_properties}.

 \begin{deluxetable*}{c|cc|cc|cc}[hbt!]
\tablecaption{\emph{Chandra} and \emph{NuSTAR's} PWN joint fit results using different pulsar properties}
\label{tab:more_fitting}
\tablewidth{0pt}
\tablehead{
\colhead{Parameter}  & \multicolumn{2}{c}{Low Boundary} &\multicolumn{2}{c}{Best Fit} &\multicolumn{2}{c}{High Boundary} }
\startdata
\multicolumn{7}{c}{Joint fit between \emph{Chandra} and \emph{NuSTAR's} PWN (3$-$10 keV)}\\
 &  (1)& (2)& (3) & (4) & (5) & (6) \\ 
$\Gamma_{PWN}$ &$1.9\pm{0.1}$ &$ 1.7\pm{0.1}$  &$2.0\pm{0.1}$ & $1.6\pm{0.1}$ &$2.1\pm{0.1}$ & $1.6
\pm{0.1}$ \\
$F_{PWN}$ &  $ (8.5 \pm{0.2}) \times 10^{-13}$&$(8.2 \pm{0.2}) \times 10^{-13}$&$(9.5\pm{0.2}) \times 10^{-13}$ &$ (8.2 \pm{0.2}) \times 10^{-13}$ & $ (10.9 \pm{0.3}) \times 10^{-13}$ & $ (9.0 \pm{0.3}) \times 10^{-13}$ \\
$\chi^{2}_{\nu}$(DOF) & 1.14 (554)  & 1.14 (554) & 1.14 (554)&1.12 (554)&1.16 (554)&1.14 (554)\\ 
\hline
\hline
\multicolumn{7}{c}{\emph{NuSTAR's} PWN (10$-$30 keV)}\\
 &  (7)& (8)& (9) & (10) & (11) & (12) \\
 $\Gamma_{PWN}$ & $8.8^{+0.0}_{-2.0}$  &$3.2 \pm{0.5}$ &$3.5\pm{0.5}$& $1.8\pm{0.2}$&$2.1\pm{0.3}$& $1.6\pm{0.2}$ \\
$F_{PWN}$ & $(3.14\pm{0.2}) \times 10^{-14}$&$(2.0 \pm{0.4}) \times 10^{-13}$&$(2.0 \pm{0.4}) \times 10^{-13}$&$(5.8 \pm{0.5}) \times 10^{-13}$&$(4.9 \pm{0.5}) \times 10^{-13}$& $(7.9\pm{0.5}) \times 10^{-13}$\\
$\chi^{2}_{\nu}$(DOF) & 1.4 (765)  & 1.09 (765) & 1.09 (765)&1.07 (765)&1.07 (765)& 1.08 (765) \\ 
\enddata
\tablecomments{This is a joint fit for \emph{Chandra} and \emph{NuSTAR's} PWN using \emph{Chandra's} derived pulsar properties in Table \ref{tab:psr_properties} after taking dust scattering into account. The un-absorbed flux, photon index were linked together across \emph{Chandra} and \emph{NuSTAR}. Errors are at the 90 \% confidence level. The reported fluxes are un-absorbed and in the units of erg ${\rm cm}^{-2}$ $s^{-1}$.}
\end{deluxetable*}

\subsubsection{Results} \label{sec:nustar_chandra_data}

\begin{center}
\begin{figure*}
\subfloat[3$-$10 keV]{\label{fig:sub_3}
\includegraphics[width=0.48\linewidth]{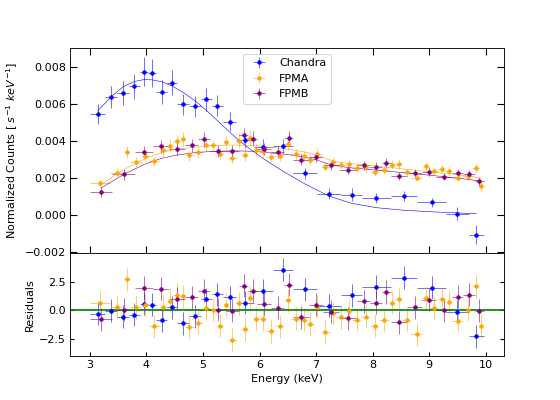}
}
\hfill
\subfloat[10$-$30 keV]{\label{fig:cdiagrams}
\includegraphics[width=0.48\linewidth]{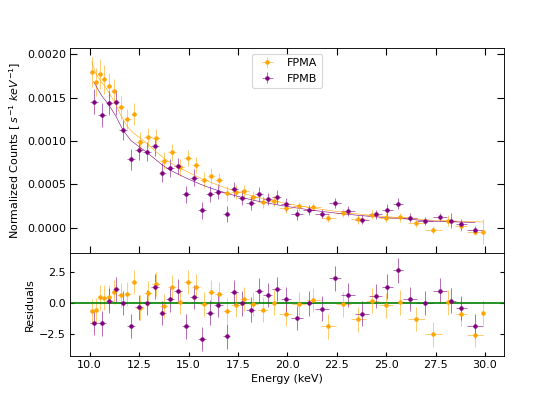}
}
\caption{A simultaneous fit of the pulsar and its PWN. (a) Result of fit (4) in Table \ref{tab:more_fitting}; \emph{Chandra} + \emph{NuSTAR} in the 3$-$10 keV energy range.  (b) Result of fit (11) in Table \ref{tab:more_fitting}; \emph{NuSTAR} in the 10$-$30 keV.}
\label{fig:pwn_fit}
\end{figure*}
\end{center}

Since \emph{Chandra} and \emph{NuSTAR} overlap in the 3$-$10 keV energy range, we jointly fit them together. For that, we used the entirety of the \emph{Chandra} selection region shown previously in Figure \ref{fig:nustar_selectiom} and used a $tbabs \times (pegpwrlw + pegpwrlw) \times constant$ model with the constant value set to 1 for the \emph{Chandra} data and 0.75 for both FPMA and FPMB to account for the cross-normalization issues between the two telescopes as explained in \S \ref{sec:cross_normalization}. The parameters of the first power law $-$representing the derived pulsar properties$-$ are shown in Table \ref{tab:psr_properties} and the joint fit results representing the PWN emission are shown in Table \ref{tab:more_fitting} and two representative joint fits are shown in Figure \ref{fig:pwn_fit}. All the fits yielded good results with an average $\chi^{2}_{\nu}$ of 1.14. It is worth mentioning that fit number (7) resulted in a much worse $\chi_{\nu}$ than all the other, therefore we decided to exclude it from our consideration. This is the result of having a very hard pulsar photon index which means most of the emission is dominated by the pulsar with little to almost no contribution from the PWN.

In the 3$-$10 keV energy range and for the same value of $N_{H}$, the PWN gets harder (decrease in $\Gamma_{PWN}$) and emits less flux as the pulsar gets softer (increase in $\Gamma_{PSR}$). For the six different combinations of the pulsar parameters we used, we get a value of $\Gamma_{PWN}$ ranging between 1.6 to 2.1 and a value of un-absorbed flux ranging from $ 8.2 \times 10^{-13}$ to $ 10.9 \times 10^{-13}$ erg ${\rm cm}^{-2}$ $s^{-1}$.

In the 10$-$30 keV energy range, we jointly fit FPMA and FPMB only, excluding the \emph{Chandra} data since it doesn't extend beyond 10 keV. The assumed pulsar emission between 10$-$30 keV is a power-law extrapolation of the pulsar emission between 3$-$10 keV, as measured by \emph{Chandra} after accounting for cross calibration offset between these two telescopes in their overlapping energy range as explained in \S \ref{sec:cross_normalization}

In this higher energy range, and for the same value of $N_{H}$, the PWN gets harder (decrease in $\Gamma_{PWN}$) and emits more flux as the pulsar gets softer (increase in $\Gamma_{PSR}$). For the six different combinations of the pulsar parameters we used, we get a value of $\Gamma_{PWN}$ ranging between 1.6 to 3.5 and a value of  un-absorbed flux ranging from $ 2.0 \times 10^{-13}$ to $ 7.9 \times 10^{-13}$ erg ${\rm cm}^{-2}$ $s^{-1}$.

In addition, for 2 out of the 3 derived pulsar parameters, we generally observe a decrease in the PWN flux and spectral softening (increase in $\Gamma_{PWN}$) as we go from the 3$-$10 keV part of the spectrum to the 10$-$30 keV. However, at an $N_{H}$ value of $ 1.9 \times 10^{23}$ ${\rm cm} ^{-2}$, the $\Gamma_{PWN}$ remains unchanged across energy bands while the flux goes down, similar to what we observed at the other two $N_{H}$ values. This decrease in flux across energy bands was observed before by \citet{2020ApJ...904...32H} in their analysis of PWN G21.5-0.9.

To summarize our spectral analysis, in the 3$-$10 keV energy band, we see a photon index ($\Gamma_{PWN}$) ranging between ($1.6  \pm{0.1}$) and ($2.1 \pm{0.1}$) and flux values ranging from $(8.2 \pm{0.2})\times 10^{-13}$ to $(10.9 \pm{0.3})\times 10^{-13}$ erg ${\rm cm}^{-2}$ $s^{-1}$. For the 10$-$30 keV energy range, we see a photon index ($\Gamma_{PWN}$) ranging between ($1.6  \pm{0.2}$) and ($3.5 \pm{0.5}$) and flux values ranging from $(2.0 \pm{0.4})\times 10^{-13}$ to $(7.9 \pm{0.5})\times 10^{-13}$ erg ${\rm cm}^{-2}$ $s^{-1}$. 

Therefore, the X-ray part of our multi-wavelength modelling will be divided into two energy ranges; soft X-ray range (3$-$10 keV) from the \emph{Chandra} \& \emph{NuSTAR} joint fit and hard X-ray range which will incorporate the the 10$-$30 keV results from \emph{NuSTAR}. The range of X-ray spectral parameters used when modelling the data is provided in Table \ref{tab:pwn_properties}.

\begin{table*}[htbp]
\caption{Summary of the X-ray properties used in modelling this source along with the values predicted by the model}
\makebox[\textwidth][c]{
\begin{tabular}{ccc}
\toprule
\toprule
Parameter & Allowed Range  & Model Prediction\\
\bottomrule
3$-$10 keV unabsorbed flux& ($8.2 - 10.9$) $\times 10^{-13}$  & 8.97 $\times 10^{-13}$  \\
3$-$10 keV photon index &  1.6 $-$ 2.1 & 2.05     \\
\hline
10$-$30 keV unabsorbed flux& $ (2.0 - 7.9) \times 10^{-13}$ & 6.95 $\times 10^{-13}$ \\
10$-$30 keV photon index& 1.6 $-$ 3.5 & 2.22 \\
\bottomrule
\end{tabular}%
}
\tablecomments{The model prediction of the X-ray spectrum was required to fall within the quoted range of allowed values. The reported fluxes are un-absorbed and in the units of erg ${\rm cm}^{-2}$ $s^{-1}$.}
\label{tab:pwn_properties}
\end{table*}%

\section{PWN Modelling} \label{sec:modelling}
PWNe are often studied to infer the properties of their progenitor supernova, associated neutron star along with its pulsar wind since these properties are hard to be measured directly. Currently, this is best achieved by using an evolutionary model of a spherical PWN inside a SNR to fit the observed dynamical and radiative properties of these PWNe (see \citealt{2017ASSL..446..161G} for a recent review of such models). 

Our work is based on the model described by \citet{2009ApJ...703.2051G} and was previously implemented to study other PWNe systems such as the Eel \citep{2022ApJ...930..148B}, Kes 75 \citep{2023ApJ...942..103S,2021ApJ...908..212G, 2014AN....335..318G}, G21.5-0.9 \citep{2020ApJ...904...32H}, G54.1+0.3 \citep{2015ApJ...807...30G} and HESS J1640-465 \citep{2014ApJ...788..155G, 2021ApJ...912..158M}.

A Metropolis Markov Chain Monte Carlo (MCMC) algorithm was implemented to identify the best fit values of our 15 model input parameters $\Theta$ outlined in Table \ref{tab:pwn_modelling} that best reproduce the observed properties of our source (see \S 3.2 of \citealt{2015ApJ...807...30G} for a full description). In addition to requiring the X-ray spectrum predicted by the model to fall within the range of X-ray spectral parameters listed in Table \ref{tab:pwn_properties}, this includes the radius of the associated SNR G338.3-0.0, $\theta_{snr}$ = $4.45^{\prime} \pm 0.5^{\prime}$ \citep{1970AuJPA..17..133S}, the radius of the PWN, $\theta_{pwn}$ = $3.3^{\prime} \pm 0.2^{\prime}$  \citep{2009ApJ...706.1269L}, the upper limit on the  660 MHz radio flux density of the PWN, $S_{660} < 690 \pm 300$ mJy \citep{2011A&A...536A..98C} and updated $\gamma$-ray fluxes listed in Table 6 of \citet{2021ApJ...912..158M}. 

\begin{table*}[htbp]
\caption{Our Best fit model parameters for HESS J1640-465 along with previous results by \citealt{2021ApJ...912..158M}}
\makebox[\textwidth][c]{
\begin{tabular}{ccc}
\toprule
\toprule
Model Parameter & Our Work &\citet{2021ApJ...912..158M} \\
\bottomrule
SN Explosion Energy $E_{sn}$ &  $1.36 \times 10^{51}$ ergs & $1.4 \times 10^{51}$ ergs\\
SN Ejecta Mass $M_{ej}$ &   10.0 $M_{\odot}$  & 11 $M_{\odot}$\\
ISM Density $n_{ism}$ & 0.03 ${cm}^{-3}$ & 0.009 ${cm}^{-3}$ \\
Pulsar Braking Index p &  3.15 (Fixed) & 3.15 (Fixed) \\
Pulsar Spindown Timescale $\tau_{sd}$ &  4.76 years & 4.15 years\\
Wind Magnetization $\eta_{B}$ &  0.07 & 0.10 \\
Min. Inejected Particle Energy $E_{min}$  & 1 GeV & 270 GeV\\
Break Inejected Particle Energy $E_{break}$ & 1 TeV & 16 TeV\\
Max. Inejected Particle Energy $E_{max}$ &  1.24 PeV & 0.3 PeV\\
Low Inejected Particle Index $p_{1}$  & 1.44 & 1.74 \\
High Injected Particle Index $p_{2}$ & 2.68 &  2.94\\
Temperature of added Photon Fields $T_{ic}$ &  354 K, 16108 K & 5.4 K, 27000 K \\
Energy Density of added photon fields $U_{ic}$ & 0.03 $\frac{\mathrm{keV}}{\mathrm{~{\rm cm}}^{3}}$, 0.17 $\frac{\mathrm{keV}}{\mathrm{~{\rm cm}}^{3}}$   &0.64 $\frac{\mathrm{keV}}{\mathrm{~{\rm cm}}^{3}}$, 1.3$\frac{\mathrm{keV}}{\mathrm{~{\rm cm}}^{3}}$  \\
Distance d & 11.9 kpc  & 12.1 kpc \\
\bottomrule
\end{tabular}%
}
\label{tab:pwn_modelling}
\end{table*}%

\begin{figure}[ht!]
\includegraphics[scale=0.5]{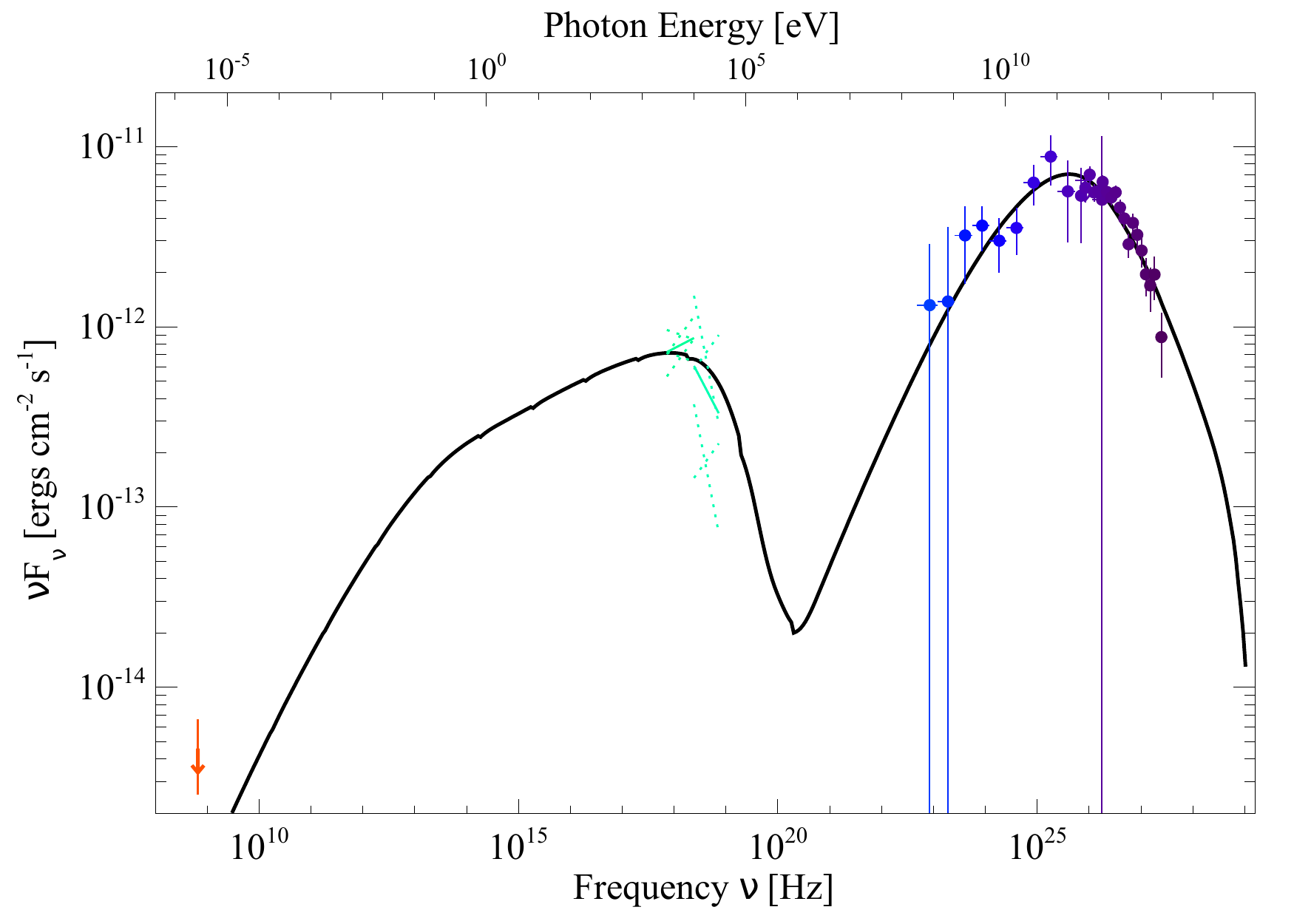}
\caption{PWN model fit to the broadband SED of HESS J1640-465. Radio upper limit (red), boundaries of X-ray data (dashed green), the average $\Gamma$ and flux in both X-ray bands inferred from these boundaries (solid-green), $\gamma$-ray data (blue \& purple). The four butterflies correspond to the different combination of the unabsorbed flux and $\Gamma$ in each X-ray band as listed in Table \ref{tab:pwn_properties}. The best fit model prediction in each X-ray band is reported in column 3 in Table \ref{tab:pwn_modelling}.}
\label{fig:hess_nustar}
\end{figure}

The true age $t_{age}$ of the system, initial spin-down luminosity $\dot{E}_{0}$ and and spin-down timescale $\tau_{sd}$ of the associated pulsar J1640-4631 are calculated by the model using its current spin-down luminosity of $\dot{E} = 4.4 \times 10^{36}$ ergs/s and its characteristic age $t_{ch}$ = 3350 years for an assumed constant braking index of p = 3.15 as measured by \citet{2016ApJ...819L..16A} :
\begin{align}
t_{age} &=\frac{2 t_{ch}}{p-1}-\tau_{sd}\\
\dot{E}_{0} &=\dot{E}\left(1+\frac{t_{\text {age }}}{\tau_{\text {sd }}}\right)^{+\frac{p+1}{p-1}}
\end{align}

\begin{table*}[htbp]
\caption{Dynamical Properties of HESS J1640-465 predicted by our evolutionary model}
\makebox[\textwidth][c]{
\begin{tabular}{ccc}
\toprule
\toprule
Model Parameter & Our Work &\citet{2021ApJ...912..158M} \\
\bottomrule
True Age $t_{age}$  & 3093 years & 3094 years \\ 
Initial Spin Period $P_{0}$ & $\approx$ 10.1 ms & $\approx$ 9.5 ms  \\
Initial Period Derivative $\dot{P}_{0}$ & $3.13 \times 10^{-11}\frac{\mathrm{s}}{\mathrm{s}}$   & $3.4 \times 10^{-11} \frac{\mathrm{s}}{\mathrm{s}}$\\
Magnetic Field Strength $B_{pwn}$ &  \SI{4.1}{\micro G}& - \\
Initial Surface Magnetic Field $B_{s0}$ & $1.8\times 10^{13}$ G&  $1.8\times 10^{13}$ G  \\
Initial Spin Down Luminosity $\dot{E}_{0}$ & $1.2 \times 10^{42} \frac{\mathrm{erg}}{\mathrm{s}}$  & $1.5 \times 10^{42} \frac{\mathrm{erg}}{\mathrm{s}}$ \\
Reverse Shock Radius $R_{rs}$  & 11.7 pc & - \\
PWN Radius $R_{pwn}$ & 12.0 pc & - \\
\bottomrule
\end{tabular}%
}
\label{tab:output_model}
\end{table*}%

\begin{deluxetable*}{cccccc}
\tablenum{10}
\tablecaption{Comparison of the properties of PSR J1640-4631 associated with HESS J1640-465 with other pulsars}
\label{tab:comparison}
\tablewidth{0pt}
\tablehead{
\colhead{Property} & \colhead{Kes 75} & \colhead{HESS J1640$-$465} & \colhead{G54.1+0.3} & \colhead{Eel}& \colhead{G21.5-0.9} }
\startdata
 $P$  & $\approx$ 329 ms$^{a}$ & $\approx$ 206 ms$^{b}$ &  $\approx$ 137 ms$^{c}$ & $\approx$ 110 ms$^{d}$&$\approx$ 61.8 ms$^{f}$    \\
 $P_{0}$ & $\approx$ 200 ms $^{a}$ & $\approx$ 10 ms  &$\approx$ 32$-$84 ms $^{c}$ & $\approx$ 82$-$90 ms $^{e}$&$\approx$ 20 ms $^{f}$\\
 $\dot{E}$ & $8.1 \times 10^{36} \frac{\mathrm{erg}}{\mathrm{s}}^{a}$ & $4.4 \times 10^{36} \frac{\mathrm{erg}}{\mathrm{s}}$ & $1.2 \times 10^{37} \frac{\mathrm{erg}}{\mathrm{s}}^{c}$  & $3.6 \times 10^{36} \frac{\mathrm{erg}}{\mathrm{s}}^{e}$ & $3.4 \times 10^{37} \frac{\mathrm{erg}}{\mathrm{s}}^{f}$ \\
 $B_{ns}$ & $4.9\times 10^{13} G^{a}$ & $1.4\times 10^{13}$ G & $1.0\times 10^{13}$ G$^{c}$ & $3.7\times 10^{12} $G$^{e}$ & $3.58\times 10^{12} $G$^{f}$\\
 $B_{LC}$& 1.45 G$^{a}$ & 0.42 G & 0.3 G$^{c}$ &0.1 G $^{e}$ & 0.1 G$^{f}$  \\
 $\tau_{sd}$ & 398 years$^{a}$& 4.78 years & 794 years$^{c}$ & (8400$-$9800) years $^{e}$  & 2900 years$^{f}$  \\
 $\eta_{B}$ &  0.115 $^{a}$ &0.07$^{b}$ &  (0.44$-$2.2)$^{c}$ $\times 10^{-3}$& (0.001$-$0.005)$^{e}$ & 0.0032 $^{f}$ \\
$E_{max}$ &1.11 PeV$^{a}$ &1.24 PeV  & 9.5 PeV$^{c}$ & (1.6$-$2.3) PeV$^{e}$  & 0.3 PeV$^{f}$  \\
\bottomrule
\enddata
\tablecomments{P: Pulsar period, $P_{0}$: The initial spin period of the pulsar, $\dot{E}$: The spin-down luminosity of the pulsar, $t_{ch}$: characteristic age of the pulsar, $t_{age}$: The true age of the pulsar, $B_{ns}$: Surface dipole magnetic field strength, $B_{LC}$: Magnetic field strength at the light cylinder,  $\tau_{sd}$: The spin-down timescale,  $\eta_{B}$: Wind magnetization and $E_{max}$: Maximum injected particle energy \\
$^{a}$ (\citealt{2023ApJ...942..103S}; \citealt{2021ApJ...908..212G})\\
$^{b}$ \citep{2014ApJ...788..155G}\\
$^{c}$ \citep{2015ApJ...807...30G}\\
$^{d}$\citep{2019MNRAS.487.1964K}\\
$^{e}$\citep{2022ApJ...930..148B}\\
$^{f}$\citep{2020ApJ...904...32H}\\
}
\end{deluxetable*}

The MCMC algorithm makes use of the maximum likelihood estimation method described in detail in \citealt{2020ApJ...904...32H}. Table \ref{tab:pwn_modelling} outlines the combination of our 15 model input parameters $\Theta$ that yielded the biggest likelihood of our MCMC runs or equivalently, the lowest $\chi^{2}$, and Table \ref{tab:output_model} shows the output dynamical properties of our model for this source. In addition, we also include previous modelling results for this source that were done by \citet{2021ApJ...912..158M} using the same model in both tables for comparison purposes and to show how our updated measurement of the X-ray spectrum can lead to different predictions by our evolutionary model. Figure \ref{fig:hess_nustar}  shows our well fitted SED for the observed properties of this source. Statistically,  $\chi^{2} = 29.9$ for 22 degrees of freedom resulting in  $\chi_{\nu}^{2} = 1.36$ where we define $\chi^{2}$ as: 
\begin{gather}
    \chi^{2}=\sum_{i=1}^{i=35}\left(\frac{\mathcal{D}_{i}-\mathcal{M}_{i}}{\sigma_{i}}\right)^{2}
\end{gather}
where $\mathcal{D}_{i}$ is the observed property, $\mathcal{M}_{i}$ is our model-predicated value for the observed property and $\sigma_{i}$ is the error on observed quantity $\mathcal{D}_{i}$. 

\section{Discussion and Conclusions } \label{sec:discussion}
For the supernova properties ($E_{sn},M_{ej}$), our values are in agreement with the values reported by \citet{2021ApJ...912..158M}, which in turn are considered to be typical values for supernova explosions (e.g \citealt{1934PhRv...46...76B}). 

For PSR J1640$-$4631, the pulsar associated with HESS J1640$-$465, we got a spin-down timescale of $\tau_{sd} \approx$ 5 years. While this value is in agreement with the value reported by \citet{2021ApJ...912..158M}, this is an extremely short timescale in comparison with a typical value of $\tau_{sd} \sim 1000$ years reported for other pulsars as shown in Table \ref{tab:comparison}. (also see \citealt{2011ApJ...741...40T}, \citealt{2014JHEAp...1...31T}, \citealt{2015ApJ...807...30G} \& \citealt{2020ApJ...904...32H}).

For young, highly energetic ( $\dot{E}$ >~ $4 \times 10^{36}$ erg $s^{-1}$ ) pulsars, \cite{2003ApJ...591..361G} observed a correlation between the X-ray spectra of pulsars and their $\dot{E}$ following this equation:
$$
\Gamma_{PSR } = 2.08-0.029 \dot{E}_{40}^{-1 / 2} {, }
$$
where $\dot{E}_{40}$ is the spin-down energy in units of $10^{40}$ erg $s^{-1}$.  By plugging in our pulsar's $\dot{E}$ = $4.4 \times 10^{36}$ erg $s^{-1}$, we get a $\Gamma_{PSR} = 0.7$. Therefore, looking back at Figure \ref{fig:scattering}, this observed correlation favors a low $N_{H}$ value of $1.0 \times 10^{23} {cm}^{-2}$ and a nearby location for the dust grain to our source. 

Using these values along with the current pulsar period of P $\approx$ 206 ms  \citep{2014ApJ...788..155G}, we can calculate the initial period $P_{0}$ and initial period derivative $\dot{P}_{0}$ using the following equation (e.g., \citealt{1973ApJ...186..249P} \& \citealt{2006ARA&A..44...17G} and the references therein): 
\begin{align}
P_{0}=P\left(1+\frac{t_{\text {age }}}{\tau_{\text {sd }}}\right)^{-\frac{1}{p-1}}
\end{align}
\begin{align}
\dot{P}(t) &=\frac{P_{0}}{\tau_{\mathrm{sd}}(p-1)}\left(1+\frac{t}{\tau_{\mathrm{sd}}}\right)^{\frac{2-\mathrm{p}}{p-1}}\\
\dot{P}_{0} &=\frac{P_{0}}{\tau_{\mathrm{sd}}(p-1)}    
\end{align}

Plugging our values in the above two equations, we get an initial period value of $P_{0} \approx 10.1$ ms, and an initial period derivative value of $\dot{P}_{0} \approx 3.13 \times 10^{-11}\frac{\mathrm{s}}{\mathrm{s}}$. Using these values for $P_{0}$ \& $\dot{P}_{0}$, we get an initial spin-down inferred surface dipole magnetic field strength of:
\begin{gather}
B_{s 0} \equiv 3.2 \times 10^{19} \sqrt{P_{0} \dot{P}_{0}} \approx 1.8 \times 10^{13} \mathrm{G}
\end{gather}
only  $\sim 30 \%$ larger than the current value of $B_{s} \approx 1.4 \times 10^{13} \mathrm{G}$, as reported by \citet{2014ApJ...788..155G}. It is also worth mentioning that while the predicted initial spin down luminosity $\dot{E}_{0} = 1.2 \times 10^{42}$ erg  $s^{-1}$ of this pulsar is significantly larger than the values derived for other systems (e.g., \citealt{2011ApJ...741...40T}; \citealt{2014JHEAp...1...31T}; \citealt{2015ApJ...807...30G}; \citealt{2020ApJ...904...32H}).

As for the magnetization of the pulsar wind, we get a value of $\eta_{B} \approx 0.1$ which is extremely higher than a typical value of $\eta_{B} \approx 2.2 \times (10^{-2}$-- $10^{-3})$ inferred from other pulsars (e.g., \citealt{2011ApJ...741...40T}; \citealt{2014JHEAp...1...31T}; \citealt{2015ApJ...807...30G}; \citealt{2020ApJ...904...32H}). This value, however, is comparable with the also high wind magnetization value predicted for Kes 75 \citep{2021ApJ...908..212G} as shown in Table \ref{tab:comparison}. It is noteworthy that both of them have an inferred dipole surface magnetic field strength $B_{ns} \gtrsim 10^{13}$ G associated with their pulsars. To put this into perspective, Table \ref{tab:comparison} shows a comparison between the pulsar parameters of PSR J1640-4631 and other systems. There seems to be a possible relation between the dipole magnetic field strength ($B_{ns}$) and the resulting wind magnetization ($\eta_{B}$); the higher the $B_{ns}$, the higher the resulting $\eta_{B}$. However, the sample size isn't big enough to draw rigorous conclusions from this, and therefore more systems need to be modelled in order to investigate how real this correlation is. It is important to mention that all these properties were inferred using the same model described by \citet{2009ApJ...703.2051G} to minimize systematic errors.

A closely related quantity is the magnetic field strength at the light cylinder $B_{LC}$, which using the current P and $\dot{P}$ listed in Table \ref{tab:comparison}, can be calculated using the following equation (check \citealt{1969ApJ...157..869G}; \citealt{2010ApJS..187..460A}):  
\begin{align}
 B_{\mathrm{LC}}=\left(\frac{3 I 8 \pi^{4} \dot{P}}{c^{3} P^{5}}\right)^{1 / 2} \approx 2.94 \times 10^{8}\left(\dot{P} P^{-5}\right)^{1 / 2} \mathrm{G}
\end{align}
 with the radius defined as ($R_{\mathrm{LC}}=c P / 2 \pi$). This will result in a value of $B_{LC} \approx 0.42 \mathrm{G}$ for our source. The reason why $B_{LC}$ is a quantity of interest is because the pulsar wind is equal to the plasma that leaves the magnetosphere at the light cylinder. Therefore, the physical properties of the wind could depend on the physical properties at the light cylinder. To investigate this, we included the $B_{LC}$ value of other systems in Table \ref{tab:comparison} for comparison purposes. Similar to $B_{NS}$, there seems to be a possible connection between the resulting value of wind magnetization $\eta_{B}$ we inferred from our model and the value of ${B_{LC}}$. However, our sample size needs to be bigger to be able to draw rigorous conclusions and check if this possible trend is real or not. 

 As for the injected particle spectrum, we get: $p_{1}$ = 1.44 and $p_{2} = 2.68$, both are harder than what was previously found by \citet{2014ApJ...788..155G} \& \cite{2021ApJ...912..158M}. In addition, our model predicts having particles with $E_{min}$ of 1 GeV, 270 times lower than what \citet{2021ApJ...912..158M} found, $E_{break}$ of 1 TeV, 16 times lower than what \citet{2021ApJ...912..158M} found. And quite importantly, particles injected in the PWNe can be accelerated up to $E_{max}$ = 1.24 PeV, which makes HESS J1640-465 to be an electronic PeVatron candidate.

As for the surrounding medium, the $\gamma$-ray emission we observed from this source is due to inverse Compton scattering of the particles inside the PWN off two photon fields on top of the Cosmic Microwave Background (CMB). The first has a temperature of 355 K, which is a typical temperature of the dust inside supernova remnants \citep{2006AJ....132.1610T} and the other is a very hot photon field with a temperature of T $\approx 16000$ K, which is a typical temperature of massive stars. The change in the particle injection spectrum described in the previous paragraph affects the properties of the colder background photon field inferred from $\gamma$-rays causing it to become hotter as we can see from Table \ref{tab:pwn_modelling}.

As previously mentioned in \S \ref{sec:intro}, PWNe are composed of highly relativistic electrons which cause them to emit electromagnetic radiation through mainly two dominant mechanisms; synchrotron radiation through the interaction of leptons with the PWN's magnetic field and inverse Compton radiation through the interaction of these leptons with ambient low-energy photons (e.g, \citealt{2006ARA&A..44...17G}). In order to fit the updated X-ray spectrum presented in this paper, a different particle injection spectrum than that reported by \citet{2021ApJ...912..158M} was needed. This invariably affected the background photon fields needed to reproduce the $\gamma$-ray measurements. As a result, the aforementioned highlighted differences between our model parameters and that of \citet{2021ApJ...912..158M}, namely the injection particle spectrum and background photon fields, can be directly attributed to our updated measurements of the X-ray spectrum of the PWN. 

Lastly, for the PWN's evolution phase, our model shows that the PWN has started to interact with the SNR's reverse shock as indicated by their radii values listed in Table \ref{tab:output_model} ($R_{rs} < R_{PWN}$). This result is in line with the asymmetrical morphology of the source that we can clearly see in Figure \ref{fig:source}. 

\section{Summary and Conclusions} \label{sec:conclusion}
We have analyzed archival X-ray (\emph{Chandra} and \emph{NuSTAR}) data for HESS J1640-465. The high value of $N_H$ along the line of sight suggests that scattering will significantly affect measurements of the pulsar spectrum. Consequently, we used the $xscat$ model incorporated in XSPEC to quantify these effects in the puslar spectrum extracted from the \emph{Chandra} data so we can propagate them to the \emph{NuSTAR} data. 

We analyzed the X-ray spectrum of the PWN in two bands; soft X-ray in the 3$-$10 keV energy range, and hard X-ray in the 10$-$30 keV energy range. Our spectral analysis showed evidence for spectral softening across bands which ultimately resulted in having range of values for the PWN properties in both energy bands. The values were reported in Table \ref{tab:pwn_properties}.

We then used a one-zone time-dependent model for the evolution of a PWN inside a SNR to reproduce the observed dynamical and multi-wavelength spectral properties of HESS J1640-465. Our model shows that PWNe can be PeVatron candidates as there are particles that can be accelerated up to $E_{max}$ = 1.24 PeV for our system. Also, Our model requires the associated pulsar to have a high initial spin down luminosity of $\dot{E}_{0} \approx 1.2 \times 10^{42}$ erg  $s^{-1}$ and a short spin-down timescale of $\tau_{sd} \approx 5$ years compared to other pulsars as shown in Table \ref{tab:comparison}. In addition, our value for the wind magnetization ( $\eta_{B}$ = 0.07 ) is higher than average and only comparable to the value obtained for Kes 75 \citep{2021ApJ...908..212G} as detailed in Table \ref{tab:comparison}. Lastly, there seems to be a possible connection between the inferred wind magnetization ($\eta_{B}$) values and the values of the surface dipole magnetic field ($B_{NS}$) and magnetic field strength at the light cylinder ($B_{LC}$). However, more systems need to be modelled in order to draw rigorous conclusions about this. 

\section*{Acknowledgement}

The contributions of JDG and SMS was supported by the National Aeronautics and Space Administration (NASA) under grant number NNX17AL74G issued through the NNH16ZDA001N Astrophysics Data Analysis Program (ADAP). JDG and SH are also supported by the NYU
Abu Dhabi Research Enhancement Fund (REF) under grant RE022. The research of JDG is also supported by NYU Abu Dhabi Grant AD022. The work of J.D.G. and S.M.S. is also supported by Tamkeen under the NYU Abu Dhabi Research Institute grant $CAP^3$.

This research has made use of data obtained from the \emph{Chandra} Data Archive and the \emph{Chandra} Source Catalog, and software provided by the \emph{Chandra} X-ray Center (CXC) in the application packages CIAO and Sherpa. It also made use of data from the \emph{NuSTAR} mission, a project led by the California institute of Technology,  managed by the Jet Propulsion Laboratory, and funded by NASA. This research made use of
the \emph{NuSTAR} Data Analysis Software (NuSTARDAS) jointly developed by the ASI Science Data Center (ASDC, Italy) and the California Institute of Technology (USA). We also made use of data and asoftware provided by the High Energy Astrophysics Science Archive Research Center (HEASARC), which is a service of the Astrophysics Science Division at NASA/GSFC. We also made use of NASA’s Astrophysics Data System Bibliographic Services.
\vspace{2.5mm}

\emph{Facilities}: CXO, NuSTAR 
\vspace{2.5mm}

\emph{Software}: Xspec \citep{1996ASPC..101...17A}, CIAO \citep{2006SPIE.6270E..1VF}, HEASoft \citep{2014ascl.soft08004N}, Sherpa (\citealt{2001SPIE.4477...76F}; \citealt{2007ASPC..376..543D}) and SAOImageDS9 (\citealt{2006ASPC..351..574J}; \citealt{2000ascl.soft03002S})

\bibliography{Manuscript}{}
\bibliographystyle{aasjournal}



\end{document}